\UseRawInputEncoding
\documentclass[12pt,prd,superscriptaddress,nofootinbib,tightenlines,floatfix,preprintnumbers]{revtex4}

\usepackage{amstext,amssymb}
\usepackage{amsmath}
\usepackage{graphicx}
\usepackage{dcolumn}
\usepackage[hyperfootnotes=false]{hyperref}
\usepackage{xspace}
\usepackage{braket}
\usepackage{color}
\usepackage{cancel}
\usepackage[normalem]{ulem}
\usepackage{slashed} 

\begin{document}
\title{Dark matter to baryon ratio from scalar triplets decay in type-II seesaw}
\author{Nimmala \surname{Narendra}}
\email{nimmalanarendra@gmail.com}
\affiliation{Theoretical Physics Division, Physical Research Laboratory, Ahmedabad-380009, India}

\author{Narendra \surname{Sahu}}
\email{nsahu@phy.iith.ac.in}
\affiliation{Department of Physics, Indian Institute of Technology Hyderabad, Kandi, Sangareddy, 502285, Telangana, India}

\author{Sujay \surname{Shil}}
\email{sujayshil1@gmail.com}
\affiliation{Institute of Physics, Sachivalaya Marg, Bhubaneswar, Odisha 751005, India}
\affiliation{Homi Bhabha National Institute, Training School Complex, Anushakti Nagar, Mumbai 400085, India}

\begin{abstract}
We propose a minimal model for the cosmic coincidence problem $\Omega_{\rm DM}/\Omega_B \sim 5$ 
and neutrino mass in a type-II seesaw scenario. We extend the standard model of particle physics with a $\rm SU(2)$ singlet 
leptonic Dirac fermion $\chi$, which represents the candidate of dark matter (DM), and two triplet scalars 
$\Delta_{1,2}$ with hierarchical masses. In the early Universe, the CP violating out-of-equilibrium decay 
of lightest $\Delta$ generates a net $B-L$ asymmetry in the visible sector (comprising of SM fields), where 
$B$ and $L$ represents the total baryon and lepton number respectively. A part of this asymmetry gets 
transferred to the dark sector (comprising of DM $\chi$) through a dimension eight operator which conserves 
$B-L$. Above the electroweak phase transition, the $B-L$ asymmetry of the visible sector gets converted to 
a net $B$-asymmetry by the $B+L$ violating sphalerons, while the $B-L$ asymmetry of the dark sector remains 
untouched which we see today as relics of DM. We show that the observed DM abundance can be explained for 
a DM mass about 8 GeV. We then introduce an additional singlet scalar field $\phi$ which 
mixes with the SM-Higgs to annihilate the symmetric component of the DM resonantly which requires the 
singlet scalar mass to be twice the DM mass, {\it i.e.} around 16 GeV, which can be searched at collider 
experiments. In our model, the active 
neutrinos also get small masses by the induced vacuum expectation value (vev) of the triplet scalars $\Delta_{1,2}$. 
In the later part of the paper we discuss all the constraints on model parameters coming from invisible Higgs 
decay, Higgs signal strength, DM direct detection and relic density of DM.
\end{abstract}


\maketitle
\newpage
\section{\label{sec:1}Introduction}
%
One of the most important aspects of beyond the standard model (SM) of particle physics is dark matter (DM) 
phenomenology. There are lots of astrophysical evidences which ensure the existence of DM~\cite{Jungman:1995df, 
Bertone:2004pz}. The prime among them are the galaxy rotation curve, gravitational lensing and the large scale 
structure of the Universe. Another important puzzle in physics is why the Universe is baryon asymmetric. The 
baryon asymmetry of the Universe is usually given in terms of $\eta=\frac{n_{B}-n_{\bar{B}}}{n_{\gamma}}=
(6.12\pm0.04)\times 10^{-10}$\,\cite{Akrami:2018vks}, where $n_{\gamma}$ is the photon number density in a 
comoving volume and is related to entropy density $s$ as $ s=7.04 \,n_\gamma$. Here $n_B$ and $n_{\bar{B}}$ 
are respectively baryon and anti baryon densities. The baryon asymmetry of the Universe is maximal 
today, {\it i.e.,} $n_{_{\overline{B}}}=0$. As a result $\eta$ can be expressed in terms of baryon abundance 
$Y_{B}$ as $\eta = 7.04 Y_B$, where $Y_{B} \equiv n_{B}/s$. The CP violation within the SM is not adequate 
to explain the present baryon asymmetry of the Universe. This is another reason why to explore the physics 
beyond the SM (BSM).

Experimentally it has been observed that the relic density of DM, given in terms of $\Omega_{\rm DM}\equiv 
\rho_{_{\rm DM}}/ \rho_{_{c}} $, is about five times larger than the relic density of baryons in the present 
Universe, {\it i.e.}, $\Omega_{\rm DM} \sim 5 \Omega_{\rm B}$, where $\Omega_{\rm DM} h^2=0.120 \pm 0.001$ 
and $\Omega_{\rm B}h^2=0.0224 \pm 0.0001$ \cite{Hinshaw:2012aka, Aghanim:2018eyx}. This 
proportionality is known to be cosmic coincidence problem. The observed DM abundance in a comoving volume can 
be given as 
\begin{equation}
Y_{\rm DM}\equiv \frac{n_{\rm DM}}{s}= 4 \times 10^{-10} \left( \frac{1 \rm GeV}{M_{\rm DM}}\right)
\left(\frac{\Omega_{\rm DM}h^2}{0.11}  \right)\,.
\end{equation} 
Thus the ratio of the abundances of DM to baryons, ${\it i.e.,} \,Y_{\rm DM}/Y_B \approx {\cal O}(1)$ for 
$M_{\rm DM} \sim 5\,{\rm GeV}$. However, the DM mass can vary from a GeV to a TeV scale depending on the amount 
of CP violation in visible and dark sectors. See for instance~\cite{Arina:2011cu,Arina:2012fb,
Arina:2012aj}\footnote{In ref.~\cite{Arina:2011cu,Arina:2012fb,Arina:2012aj} an inert fermion doublet $\psi$ was 
introduced which symmetrically couples to $\Delta$. If $\psi$ is odd under a remnant $Z_{2}$ discrete symmetry, 
then the neutral component of $\psi$ can be a candidate of inelastic asymmetric DM.}. The idea of asymmetric DM 
is similar to the baryon asymmetry of the present Universe, {\it i.e.,} an asymmetry in DM particle over its 
anti-particle \cite{Arina:2011cu,Arina:2012fb,Arina:2012aj,Nussinov:1985xr, Griest:1986yu,Chivukula:1989qb,Dodelson:1991iv,
Barr:1991qn,Kaplan:1991ah,Dreiner:1992vm,Inui:1993wv,Thomas:1995ze,Cosme:2005sb,Tytgat:2006wy,Kitano:2004sv,
Agashe:2004bm,Farrar:2005zd,Kitano:2008tk,Nardi:2008ix,An:2009vq,Cohen:2009fz,Shelton:2010ta,Davoudiasl:2010am,
Haba:2010bm,Buckley:2010ui,Gu:2010ft,Blennow:2010qp,McDonald:2011zza,Hall:2010jx,Heckman:2011sw,Frandsen:2011kt,
Tulin:2012re,Kohri:2009ka,Kohri:2009yn,Kohri:2013sva,Graesser:2011wi,Hooper:2004dc,Iminniyaz:2011yp,Haba:2011uz,
Kang:2016nwp,Blum:2012nf,Fujii:2002aj,Banks:2006xr,Dulaney:2010dj,Cohen:2010kn,Dutta:2010va,Falkowski:2011xh,
MarchRussell:2011fi,Graesser:2011vj,Kamada:2012ht,Walker:2012ka,Feldstein:2010xe,MarchRussell:2012hi,Cai:2009ia,
An:2010kc,Kouvaris:2011fi,Buckley:2011kk,Chang:2011xn,Profumo:2011jt,Davoudiasl:2011fj,Masina:2011hu,Lin:2011gj,
Buckley:2011ye,Davoudiasl:2012uw,Okada:2012rm,Hugle:2018qbw,Petraki:2013wwa,
Narendra:2017uxl,Zurek:2013wia,Bell:2011tn,Petraki:2011mv,vonHarling:2012yn,Lonsdale:2018xwd,Novikov:2016hrc,
Novikov:2016fzd}.
\begin{figure}[h!]
				\centering
				\includegraphics[width = 80mm]{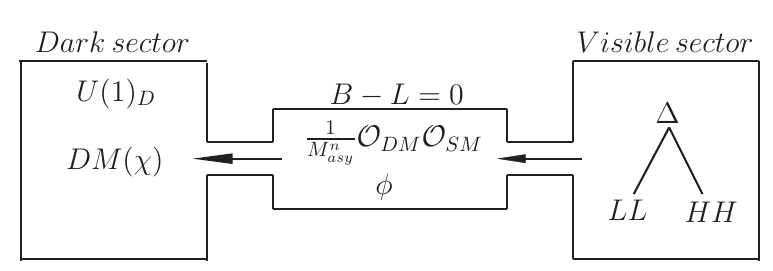}
                \caption{\footnotesize{Pictorial presentation of thermal contact between the dark sector and visible sector via scalar portal as well as higher dimension operator, which conserves $B-L$ symmetry and are in thermal equilibrium above sphaleron decoupling temperature.}}
                \label{asydm_cartoon}
\end{figure}

In this article we propose a model to explain the cosmic coincidence problem: $\Omega_{\rm DM}/\Omega_B \sim 5$ 
in a simple extension of the SM. We introduce two $\rm SU(2)$ triplet scalar fields ($\Delta_{1,2}$), which decay 
to the SM lepton doublet and Higgs field \cite{Arina:2012aj,Arina:2011cu,Arina:2012fb,Agrawal:2018pci,Ma:1998dx,Hambye:2000ui,Akeroyd:2007zv, Sierra:2014tqa} in the early Universe as pictorially shown in Fig.\,\ref{asydm_cartoon}. The triplets decay satisfy 
all the Sakharov conditions\,\cite{Sakharov:1967dj,Chen:2007fv,Fong:2013wr,Kolb:1990vq} to give the lepton asymmetry in the 
SM sector\,\cite{Fukugita:1986hr}. An extra scalar $\phi$ and a leptonic Dirac fermion $\chi$ have been introduced 
in the model. The leptonic Dirac fermion $\chi$ is charged under a global $\rm U(1)_{D}$ symmetry which provides 
stability to the DM and also forbids the Majorana mass term of the $\chi$, the term which can spoil the asymmetry 
in the dark sector. The asymmetry created in the leptonic sector transfers to the DM sector through a dimension-8 operator 
${\cal O}_8=\frac{\bar{\chi}^{2}(LH)^{2}}{M_{asy}^{4}}$\cite{Kaplan:2009ag,Feng:2012jn,Ibe:2011hq,Feng:2013wn}, which 
conserves $B-L$ symmetry. Since the operator ${\cal O}_8$ conserves $B-L$, it redistributes the $B-L$ asymmetry, produced by 
$\Delta$-decay, between SM leptons (visible sector) and $\chi$ (dark sector). Above the electroweak phase transitions, the $B+L$ 
violating sphaleron processes convert the $B-L$ asymmetry of the visible sector to a net baryon (B) asymmetry that we 
observe today, while the $B-L$ asymmetry of the dark sector remains untouched which we see today as DM abundance. We show 
that correct relic density of DM requires its mass to be around 8 GeV. The symmetric component of $\chi$ gets 
annihilated resonantly via $\phi- H$ mixing, which in turns requires the singlet scalar $\phi$ mass to be around twice 
the DM mass. Note that the dimension eight operator ${\cal O}_8$ softly breaks the $U(1)_{D}$ global symmetry to a remnant $Z_{2}$ 
symmetry under which $\chi$ is odd and all other particles are even. In this way, we ensure the stability to DM as well 
as the theory escapes from having a Goldstone boson. 

The advantages of considering an asymmetric DM in this paper are as follows: (1) The hitherto null detection of WIMP DM 
(with a typical mass $\mathcal{O}(100)$ GeV and having a weak interaction cross-section) at leading direct DM search experiments like 
XENON1T~\cite{Xenon1T} attracts many other experiments probing DM at low mass regime, where XENON1T is insensitive. The 
prime among them are CRESST-II~\cite{Angloher:2015ewa}, EDELWEISS~\cite{Arnaud:2017usi}, CDMS-II~\cite{Agnese:2013rvf}, 
CoGeNT~\cite{Aalseth:2012if}, DAMIC~\cite{Barreto:2011zu}, SuperCDMS low mass~\cite{Agnese:2014aze}, CDMSlite with 
Ge~\cite{Agnese:2015nto}, PandaX-II~\cite{Tan:2016zwf}, ZEPLIN-III~\cite{Akimov:2011tj}, DarkSide-50~\cite{Agnes:2014bvk} 
{\it etc.}. These experiments will shed light on a low mass DM, typically mass less than 10 GeV as we have considered in 
this paper. (2) The other advantage is the search of a light scalar is under active consideration at LHC~\cite{cms-1,cms-2,cms-3,
cms-4,atlas-1,atlas-2}. Therefore, there exist a fair chance that future data can shed light to our model. Since the DM and 
singlet scalar masses are correlated in our case, the model thus in principle can be probed in future.   
  
The paper is organized as follows: In Sec.\,\ref{Model}\,, we introduce the model part. Sec.\,\ref{neutino mass}\, 
is devoted to explain the neutrino masses in a type-II seesaw framework. We discuss the generation of lepton 
asymmetry in the visible sector and its transfer to a dark sector in Sec.\,\ref{trip_scalar_lep}\,. In 
Sec.\,\ref{symmetric_ann}, we show the condition for depletion of symmetric component of the DM. The constraints 
from invisible Higgs decay, Higgs signal strength, the relic abundance of DM and its direct detection on the model 
parameters are discussed in Sec.\,\ref{signal_strength}, \ref{invisible}, \ref{dd}, respectively. In 
Sec.\,\ref{conclusion}\, we conclude. 

\section{The model} \label{Model}
Here we extend the SM of particle physics with a $ U(1)_{D}$ global symmetry. The additional particle 
content that we introduce to the SM are: two triplet scalars $\Delta_{1,2}$, a singlet leptonic Dirac 
fermion $\chi$ and a $SU(2)_{L}$ singlet scalar $\phi$. The latter mixes with the SM-Higgs $H$ 
and provides interesting low energy phenomenology as we discuss in following sections. The $U(1)_{D}$ global 
symmetry is softly broken by the dimension eight operator\footnote{We have given a 
viable origin of dimension-8 operator in the appendix-B.} ${\cal O}_8=\frac{\bar{\chi}^{2}(LH)^{2}}{M_{asy}^{4}}$ 
to a remnant $Z_2$ symmetry under which the leptonic Dirac fermion $\chi$ is odd while all other particles 
are even. As a result $\chi$ is stable and represents a candidate of DM. The details of dimension 
eight operator ${\cal O}_8$ will be discussed in Sec.\,\ref{trip_scalar_lep}. The particle content of the model
and the corresponding quantum numbers are given in the Table~\ref{tab}\,. 

\begin{table}[htbp]
\centering
\begin{tabular}{|c|c|c|c|c|}
\hline
Fields &  SU(3)$_c$ & SU(2)$_L$ & U(1)$_Y$ & $U(1)_{D}$  \\
\hline
$\Delta$ & 1 & 3 & +1 & 0  \\
$\phi$ & 1 & 1 & 0 & 0  \\
$\chi$ & 1 & 1 & 0 & -1  \\
\hline
\end{tabular}
\caption{Quantum numbers of the new particles under the imposed symmetry.}
\label{tab}
\end{table}

The Lagrangian of our model can be written as:
\begin{eqnarray}
\mathcal{L} &\supset & \overline{\chi} i \gamma^{\mu} \partial_{\mu} \chi + (\partial_{\mu} \phi)(\partial^{\mu} \phi) - M_{\chi}\overline{\chi} \chi - \lambda_{\rm DM}
\overline{\chi} \chi \phi -\lambda \overline{(L^{c})}i\tau^{2}\Delta L -  V(H, \phi)\,,
\label{Lagrangian}
\end{eqnarray}
where
\begin{eqnarray}\label{potential}
V(H,\phi) &=& -\mu_{H}^{2} H^{\dagger}H + \lambda_{H}(H^{\dagger}H)^{2}+ M_{\phi}^{2}\phi^{2}
+ \lambda_{\phi}\phi^{4} + M_{\Delta}^{2} \Delta^{\dagger}{\Delta} + \lambda_{\Delta}(\Delta^{\dagger}\Delta)^{2} \nonumber\\
& + & [\mu \overline{(H^{c})}i\tau^{2}\Delta^{\dagger} H + h.c.] +  \rho_{1} \phi (H^{\dagger}H) + \lambda_{H\phi}(H^{\dagger}H)\phi^{2}\nonumber\\
&+& \rho_{2} \phi (\Delta^{\dagger}\Delta) + \lambda_{H \Delta}(H^{\dagger}H)(\Delta^{\dagger}\Delta) 
+ \lambda_{\Delta \phi} (\Delta^{\dagger} \Delta) \phi^{2}\,.
\end{eqnarray}
Here we assume the mass of $\Delta$ to be super heavy as we need to explain the small neutrino 
masses(see Sec.\ref{neutino mass})\,\cite{Ma:1998dx}. Therefore $\Delta$ does not play any role in the 
low energy phenomenology. The necessary and sufficient conditions for the vacuum stability of the 
potential are~\citep{Kannike:2012pe,Arhrib:2011uy}:
\begin{eqnarray}
\lambda_{H} \geq 0 , \lambda_{\phi} \geq 0, \lambda_{\Delta} \geq 0 \,,\\
\widetilde{\lambda_{1}}=\frac{1}{2}\lambda_{H\Delta}+\sqrt{\lambda_{H}\lambda_{\Delta}} \geq 0\,, \\
\widetilde{\lambda_{2}}=\frac{1}{2}\lambda_{H\phi}+\sqrt{\lambda_{H}\lambda_{\phi}} \geq 0\,, \\
\widetilde{\lambda_{3}}=\frac{1}{2}\lambda_{\Delta \phi}+\sqrt{\lambda_{\Delta}\lambda_{\phi}} \geq 0 \,,\\
\sqrt{\lambda_{H}\lambda_{\Delta}\lambda_{\phi}}+\frac{1}{2}\lambda_{H\Delta}\sqrt{\lambda_{\phi}}
+\frac{1}{2}\lambda_{H\phi}\sqrt{\lambda_{\Delta}}+\frac{1}{2}\lambda_{\Delta \phi}\sqrt{\lambda_{H}} 
+ \sqrt{2\widetilde{\lambda_{1}}\widetilde{\lambda_{2}}\widetilde{\lambda_{3}}} \geq 0\,.
\end{eqnarray}

We assume $M_\phi^2$ to be positive, so that $\phi$ does not acquire any direct vev. However, the electroweak 
phase transition induces a non-zero vev to $\phi$ due to the trilinear term $\rho_{1}\phi (H^{\dagger}H)$ as given 
in Eq.\,\ref{potential}. We assume that $\langle \phi \rangle =u << v$, where $v$ is the SM-Higgs vev. The 
quantum fluctuation of these fields around the minimum can be given as:

\begin{equation}\label{eqn2.8}
H=\begin{pmatrix} 0 \\ \frac{v+h}{\sqrt{2}} \end{pmatrix}\,\,\, {\rm and} \,\,\, \phi=\frac{u+\tilde{\phi}}{\sqrt{2}}.
\end{equation}

\begin{figure}[h!]
				\centering
				\includegraphics[width = 70mm]{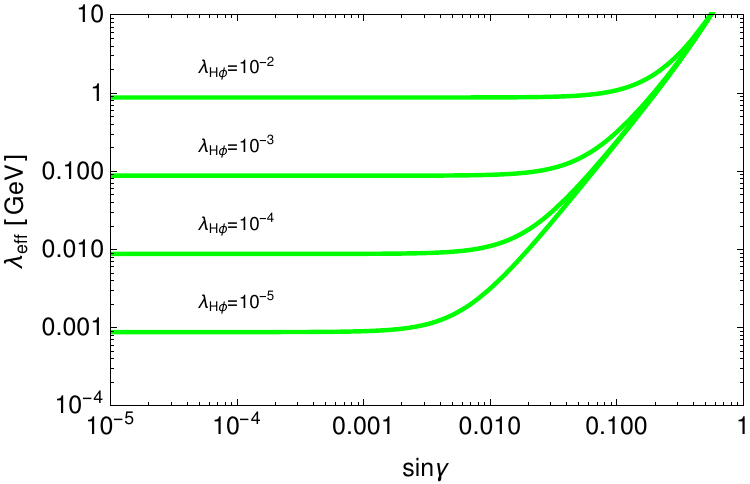}
				\includegraphics[width = 48mm]{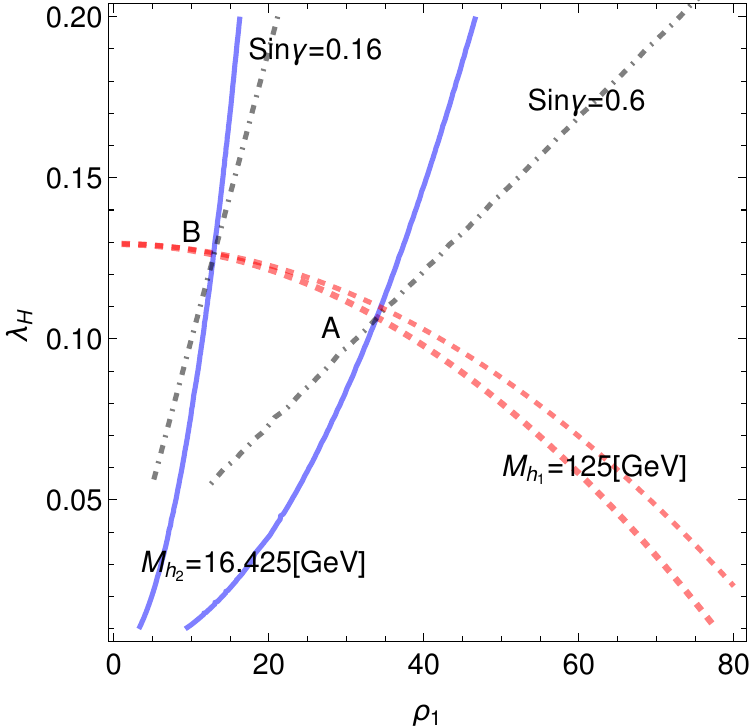}
                 \caption{\footnotesize{Left: $\lambda_{eff}$ (GeV) as a function of $\sin\gamma$. Right: Simultaneous solution to the masses 
of $h_{1}$ and $h_{2}$ for different set of values of $\sin\gamma$} and $\lambda_{H\phi}$.}
\label{singamma}
\end{figure}

Minimizing the scalar potential\,\ref{potential}, we get the vevs

\begin{equation}
v=\sqrt{\frac{\mu_{H}^{2}-\frac{1}{2}\lambda_{H\phi}u^{2}-\frac{1}{\sqrt{2}}\rho_{_1}u}{\lambda_{H}}}\,,
\end{equation}
and
\begin{equation}
u=-\frac{\rho_{_1} v^{2}}{2\sqrt{2}(M_{\phi}^{2}+\frac{1}{2}\lambda_{H\phi}v^{2})}\,.
\end{equation}
Thus from the above equation we see that as $\rho_{_1} \rightarrow 0$, the induced vev $u \rightarrow 0$. 
We note that the non zero vev of $\phi$ does not affect the discussion in the following 
sections~\footnote{Non-zero vev of $\phi$ only modifies the mass of the dark matter which ultimately fixed by 
cosmic co-incidence condition but all the other discussions will be the same. Moreover, the non-zero vev of 
$\phi$ does not affect any low energy phenomenological calculations such as dark matter annihilation 
and direct detection, bound from the Higgs measurement {\it etc}. Therefore, we set the vev of $\phi$ to be 
zero for simplicity.}, hence we set it to be zero ({\it i.e.}, 
$\langle \phi \rangle=0$) for simplicity. As a result Eq. \ref{eqn2.8} can be rewritten as: 
\begin{equation}
H=\begin{pmatrix} 0 \\ \frac{v+h}{\sqrt{2}} \end{pmatrix}\,\,\, {\rm and} \,\,\, \phi= \tilde{\phi}/\sqrt{2}\,.
\end{equation}
Due to $\tilde{\phi}-h$ mixing we get the mass matrix:  
\begin{equation}
\begin{pmatrix} 2 \lambda_H v^2  & \frac{\rho_{_1} v}{\sqrt{2}}\\
\frac{\rho_{_1} v}{\sqrt{2}} & M_\phi^2+ \frac{\lambda_{H\phi}}{2}v^2 \end{pmatrix}\,.
\label{scalar mass matrix}
\end{equation}
After diagonalising the mass matrix, given by Eq.\,\ref{scalar mass matrix}, we get the two mass 
eigenstates: 
\begin{eqnarray}
h_1 &=& h \cos \gamma + \tilde{\phi} \sin \gamma \nonumber\\
h_2 &=& - h \sin \gamma + \tilde{\phi} \cos \gamma\,.
\label{new mass eigen states}
\end{eqnarray} 
We identify $h_{1}$ to be the SM-like Higgs with mass $M_{h_{1}}=125.18 \,{\rm GeV}$, while $h_{2}$ remains as 
the second Higgs whose mass is going to be determined from the required phenomenology. In particular, we 
determine $h_{2}$ mass from relic abundance requirement. In Sec.\,\ref{symmetric_ann} we obtain it's mass 
as per the requirement of depletion of the symmetric component of the DM to be $M_{h_2} \approx 2 
M_\chi\approx 16 \,{\rm GeV}$. The $\tilde{\phi}-h$ mixing obtained from Eq.\,\ref{scalar mass matrix} is given by:
\begin{equation}
\sin \gamma \approx \frac{\rho_{_1} v/\sqrt{2}}{2\lambda_H v^2 - M_\phi^2 -\frac{\lambda_{H\phi} v^2}{2}}\,.
\label{higgs-mixing}
\end{equation}
From Eqs.\,\ref{scalar mass matrix} and \ref{higgs-mixing} we see that $M_{h_1}$, $M_{h_2}$ and $\sin \gamma$ primarily 
depend on $\rho_1$, $\lambda_H$, $\lambda_{H\phi}$. Without loss of generality we set $M_\phi=0$.  In the right panel 
of Fig.\,\ref{singamma} we show the contours of $M_{h_1}=125$ GeV (dashed lines) and $M_{h_2}=16.425$ GeV 
(solid lines) and $\sin \gamma= 0.16, 0.6$ (dot-dashed lines) in the plane of $\lambda_H$ versus $\rho_1$ for $\lambda_{H\phi}=0.01$ 
(meeting at point B) and 0.1 (meeting at point A).  Thus we see that a large range of  $\sin \gamma$ is allowed to explain 
simultaneously the masses of $h_{1}$ and $h_{2}$. However, we shall show in sections \ref{signal_strength}, \ref{invisible} and \ref{dd} 
that $\sin \gamma \gtrsim 0.2$ is not allowed.

Now we write the effective coupling of $h_{1}h_{2}h_{2}$-vertex which is relevant for the collider signature of $h_2$, studied 
in Sec.\,\ref{collider-signature}. Form Eq.\,\ref{potential}, the $h_{1}h_{2}h_{2}$-vertex can be derived 
as: 
\begin{eqnarray}
\lambda_{\rm eff} &=& 3 \lambda_H \, v \cos \gamma \sin^2 \gamma + \frac{\lambda_{H\phi}}{2} v\cos^3 \gamma +\frac{\rho_{_1}}{2\sqrt{2}} \sin^3 \gamma\nonumber\\
&-& \frac{\rho_{_1}}{\sqrt{2}} \sin\gamma \cos^2\gamma -\lambda_{H\phi} v \sin^2 \gamma \,\cos \gamma \,. 
\end{eqnarray}
In the left-panel of Fig.\,\ref{singamma} we show $\lambda_{\rm eff}$ as a function of $\sin \gamma$ for different values of $\lambda_{H\phi}$. One can see that $\lambda_{\rm eff}$ is almost independent of $\lambda_{H\phi}$ for $\sin \gamma \gtrsim 0.1$\,. We will use it to obtain the production cross-section of $h_2$ in Sec.\,\ref{collider-signature}.

\section{Neutrino masses}\label{neutino mass}
This model also explains the sub-eV neutrino masses of light neutrinos through the type-II seesaw~\cite{Konetschny:1977bn,
Cheng:1980qt,Lazarides:1980nt,Schechter:1980gr,Mohapatra:1980yp,Perez:2008ha}. The relevant terms in the 
Lagrangian (\ref{Lagrangian}) are given as:
\begin{equation}
\mathcal{L} \supset  M_{\Delta}^{2} \Delta^{\dagger} \Delta + \lambda \overline{(L^{c})}i\tau^{2}\Delta L 
+ \mu \overline{(H^{c})}i\tau^{2}\Delta^{\dagger} H + h.c.\,.
\end{equation}
Since we assume $M_{\Delta}^2 > 0$, so $\Delta$ does not acquire any direct vev. However, after electroweak 
phase transition, the trilinear term $\mu \Delta^\dagger HH$ induces a non-zero vev to $\Delta$ as:
\begin{equation}\label{delta_vev}
\langle \Delta \rangle \equiv w \simeq  \frac{-\mu v^{2}}{M_{\Delta}^{2}}\,,
\end{equation}
Where $v=\langle H \rangle = 246$ GeV. This can be verified by minimising the scalar potential, Eq.\,\ref{potential}. 
Note that the sign of $\langle\Delta \rangle$ in Eq.\,\ref{delta_vev} is not important\footnote{See for example: \,\cite{Agrawal:2018pci,Ma:1998dx,Hambye:2000ui,Akeroyd:2007zv,Sierra:2014tqa} 
for negative sign of $\langle \Delta \rangle$ and\,\cite{Cai:2017mow,Perez:2008ha} for the positive sign of 
$\langle \Delta \rangle$.} as it gives an overall phase, $e^{i \pi}$ (which is not a physical quantity), to 
the neutrino mass through the relation
\begin{equation}
(M_{\nu})_{\alpha \beta}= \lambda_{\alpha \beta} \langle \Delta \rangle = \left(\lambda_{\alpha \beta} \,
\frac{-\mu v^{2}}{M_{\Delta}^{2}} \right)\,.
\end{equation}
For $\lambda \approx \mathcal{O}(1)$, to get observed light neutrino masses, we choose $\mu \sim M_{\Delta} 
\sim 10^{14}$ GeV. Note that the electroweak $\rho$ parameter constrains the vev of $\Delta$ to satisfy the 
requirement of
\begin{equation}
\rho \equiv \frac{M_{w}^{2}}{M_{Z}^{2}\cos \theta_{w}}=\frac{1 + 2 x^{2}}{1 + 4 x^{2}} \approx 1
\end{equation}
where $x= w / v$. This implies, $ w  < \mathcal{O}(1)$ GeV. 

\section{Triplet scalar leptogenesis and asymmetric DM}\label{trip_scalar_lep}\label{lep_dm}
 In the early Universe the triplet scalars are assumed to be in thermal equilibrium at a temperature 
above their mass scales. As the Universe expands and the temperature falls below the mass 
scale of a triplet scalar, the latter goes out of thermal equilibrium and decays through the 
processes: $\Delta \to L L $ and $\Delta \to  H H $. These decay channels combinely violate $B-L$ by two units and 
hence lead to leptogenesis as we discuss below. The decay rate of $\Delta$ is given by
\begin{equation}
\Gamma_{\Delta}=\frac{1}{8\pi}(|\lambda|^{2}+ \frac{\mu^{2}}{M_{\Delta}^{2}})M_{\Delta}\,.
\end{equation} 
The out of equilibrium condition of $\Delta$ is set by comparing $\Gamma_{\Delta}$ with the Hubble expansion 
parameter $H=1.67 g_*^{1/2}T^2/M_{\rm Pl}$ at $T \sim M_{\Delta}$, and it is given by
\begin{equation} 
|\lambda| \lesssim |\sqrt{1.67\times 8 \pi \sqrt{g_{*}}(M_{\Delta}/M_{\rm Pl})-(\mu^{2}/M_{\Delta}^{2})}\,|\,.
\end{equation} 
Here the $g_{*}$ and $M_{\rm Pl}$ are the total degrees of freedom and the Planck mass respectively. For 
$\mu \leq M_{\Delta} \simeq 10^{14}$ GeV, we get $|\lambda| \lesssim \mathcal{O}(1)$. The decoupling epoch 
can be different for different mass scale of triplet scalars. 

To get the CP asymmetry we need at least two copies of triplet scalars. In presence of these triplet scalars 
interactions, the diagonal mass $M_{\Delta}^{2}$ in Eq.\,\ref{potential} can be replaced by \cite{Ma:1998dx},
\begin{equation}
\frac{1}{2}\Delta_{a}^{\dagger}(M_{+}^{2})_{ab}\Delta_{b}+\frac{1}{2}(\Delta_{a}^{*})^{\dagger}(M_{-}^{2})_{ab}\Delta_{b}^{*}.
\label{triplet_int}
\end{equation} 
The trilinear couplings $\mu \overline{(H^{c})}i\tau^{2}\Delta^{\dagger} H + h.c.$ in Eq.\,\ref{potential}\, then 
becomes $\sum\limits_{a=1,2}\mu_{a} \overline{(H^{c})}i\tau^{2}\Delta^{\dagger} H + h.c.$.

The mass matrix in Eq.\,\ref{triplet_int} can be given as:
\begin{equation}
M^{2}_{\pm}= 
\begin{pmatrix} M_{1}^{2}-i C_{11} & -i C_{12}^{\pm}\cr\\
-i C_{21}^{\pm} & M_{2}^{2}-i C_{22} \end{pmatrix}\,,
\label{heavy_mass_matrix}
\end{equation}
where
$C_{ab}^{+}=\Gamma_{ab}M_{b}$, $C_{ab}^{-}=\Gamma^{*}_{ab}M_{b}$ and $C_{aa}=\Gamma_{aa}M_{a}$ with
\begin{equation}
\Gamma_{ab}M_{b}=\frac{1}{8\pi}\left(\mu_{1a}\mu_{2b}^{*}+M_{a}M_{b} \sum\limits_{\alpha \beta} \lambda_{1 \alpha \beta} \lambda_{1 \alpha \beta}^{*}\right).
\end{equation}

Diagonalizing the above mass matrix Eq.\,\ref{heavy_mass_matrix}, we get two mass eigenvalues $M_{1}$ and $M_{2}$ 
corresponding to the two eigenstates $\xi_1^{\pm}$ and $\xi_2^{\pm}$. Note that the mass eigenstates $\xi_1^+$ 
and $\xi_1^-$ are not CP conjugate states of each other even though they are degenerate. Similarly $\xi_2^+$ and $\xi_2^-$ 
are not CP conjugate states, but they are degenerate. Therefore, the decay of these states can generate lepton asymmetry. 
We assume $M_1 \ll M_2$. As a result the asymmetry created by $\xi_2^\pm$ decay will be erased by the lepton 
number violating process mediated by $\xi_{1}^\pm$ at the temperature below the mass of $\xi_{2}^\pm$ and finally we are left 
with only $\xi_{1}^\pm$ which generate the lepton asymmetry of the Universe.

\begin{figure}[h!]
				\centering
				\includegraphics[width = 80mm]{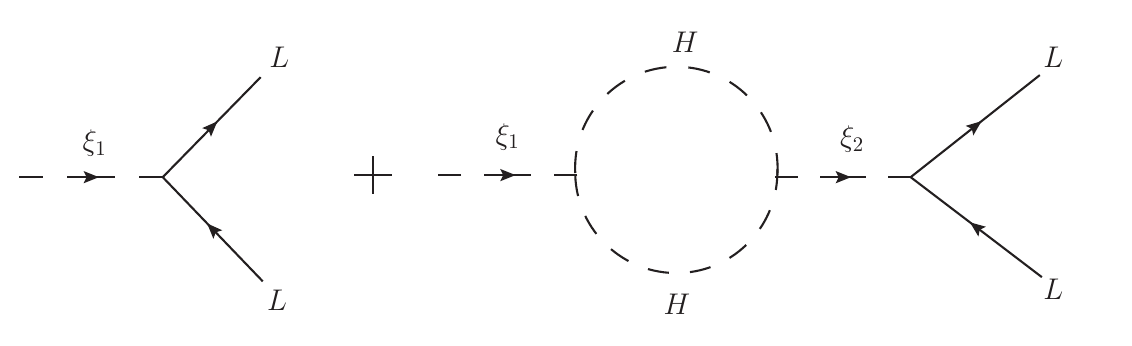}
                 \caption{\footnotesize{The tree and one loop self energy diagrams in decay of $\xi_1$.}}
\label{self_energy}
\end{figure}

The interference between tree and one loop self energy diagram of the process $\xi_{1}$ to $L L^{c}$ gives the CP asymmetry 
as shown in Fig.\,\ref{self_energy}\,.
The asymmetry $\epsilon_{L}$ is estimated to be\,\cite{Arina:2011cu} 
\begin{eqnarray}
\epsilon_{L} &=& \frac{\Gamma(\xi_{1}\rightarrow L L^{c}) - \Gamma (\xi_{1}^{c} \rightarrow L^{c} L)}{\Gamma_{1}}\nonumber\\
               &= & \frac{ {\rm Im} \left( \mu_{1} \mu_{2}^{*} \sum\limits_{\alpha \beta} \lambda_{1 \alpha \beta} \lambda_{1 \alpha \beta}^{*} \right) } { 8\pi^{2} (M_{2}^{2}-M_{1}^{2}) } \left( \frac{M_{1}}{\Gamma_{1}} \right) \,.
\end{eqnarray}
where we assume $M_1 \ll M_2$. As a result, below the mass scale of $\xi_{1}$, we get a net $B-L$ 
asymmetry~\cite{Buchmuller:1999cu,Buchmuller:2004nz,Giudice:2003jh,Narendra:2018vfw}:
\begin{equation}
(n_{B-L})_{total} = \epsilon_{L}\, \kappa s \times \frac{n_{\xi_{1}}^{eq} (T \rightarrow \infty )}{s}\,,
\label{nb-ltotal_final}
\end{equation}
where $(n_{\xi_{1}}^{eq}/s)(T \rightarrow \infty)=135 \zeta(3)/(4 \pi^{4} g_{*})$ is the relativistic equilibrium 
abundance of $\xi_{1}$. The $\kappa$ is a washout factor, arises via inverse decay and scattering processes and  
$s=(2\pi^2/45)g_* T^3$ is the entropy density. Depending on the strength of Yukawa coupling, the value of $\kappa$ 
can vary between 0 to 1. By solving the Boltzmann equations one can precisely calculate the value of 
$\kappa$~\cite{Arina:2011cu, Arina:2012fb}.
The evolution of $X_{\xi_{1}}=n_{\xi_{1}}/s$ is given by the Boltzmann equation
\begin{equation}\label{xi-density}
\frac{dX_{\xi_{1}}}{d z}= -\frac{\Gamma_{D}}{z H(z)} \left( X_{\xi_{1}}-X_{\xi_{1}}^{eq} \right) - \frac{\Gamma_{a}}{z H(z)} 
\left( \frac{X_{\xi_{1}}^{2}-X_{\xi_{1}}^{eq \,2}}{X_{\xi_{1}}^{eq}} \right)\,.
\end{equation}
Here the temperature dependent decay rate is given by:
\begin{equation}
\Gamma_{D}=\Gamma_{1} \frac{K_{1}(z)}{K_{2}(z)},\,\, {\rm where}\,\,\Gamma_{1}=\frac{1}{8\pi}\frac{|m_{\nu}| M_{1}^{2}}{v^{2}\sqrt{B_{L}B_{H}}}, \,\,\,{\rm and}\,\,\,\,\, H(z)=\frac{H(T=M_{1})}{z^{2}}. 
\label{Gamma_D}
\end{equation}
The $B_{L,\,H}$ are the branching ratios of $\xi_{1}\rightarrow LL$ and $\xi_{1}\rightarrow H H$ respectively such that $B_L + B_H =1$. 
The $K_{1,2}$ are the modified Bessel functions. In Eq.\,\ref{xi-density}, $\Gamma_{a}=\gamma_{a}/n_{\xi_{1}}^{eq}$, where the $\gamma_{a}$'s 
are the scattering densities for various processes and are given by:
\begin{eqnarray}\label{scattering-densities}
\gamma(\xi_{1}^{+}\xi_{1}^{-} \rightarrow \bar{f}f) &=& \frac{M_{1}^{4}(6 g^{4}+5 g'^{\,4})}{128 \pi^{5} z} \int_{x_{min}}^{\infty} dx \sqrt{x} K_{1}(z\sqrt{x})r^{3}, \nonumber\\
\gamma(\xi_{1}^{+}\xi_{1}^{-} \rightarrow H^{\dagger}H) &=& \frac{M_{1}^{4}(g^{4}+g'^{\,4}/2)}{512 \pi^{5} z} \int_{x_{min}}^{\infty} dx \sqrt{x} K_{1}(z\sqrt{x})r^{3}, \nonumber\\
\gamma(\xi_{1}^{+}\xi_{1}^{-} \rightarrow W^{a}W^{b}) &=& \frac{M_{1}^{4} g^{4}}{64 \pi^{5} z} \int_{x_{min}}^{\infty} dx \sqrt{x} K_{1}(z\sqrt{x})\left[r(5+34/x)-\frac{24}{x^{2}}(x-1)\ln\left(\frac{1+r}{1-r}\right) \right], \nonumber\\
\gamma(\xi_{1}^{+}\xi_{1}^{-} \rightarrow B B) &=& \frac{3 M_{1}^{4} g'^{4}}{128 \pi^{5} z} \int_{x_{min}}^{\infty} dx \sqrt{x} K_{1}(z\sqrt{x})\left[r(1+4/x)-\frac{4}{x^{2}}(x-2)\ln\left(\frac{1+r}{1-r}\right) \right]  \nonumber\\
\end{eqnarray}
where $z=M_1/T$, $r=\sqrt{1-4/x}$ and $x=\hat{s}/M_{1}^{2}$ with $\hat{s}$ is the Mandelstam variable, the center of mass energy. In 
Eq.\,\ref{scattering-densities}, $g, g'$ are the gauge couplings corresponding to $SU(2)_{L}$ and $U(1)_{Y}$ respectively. The $n_{\xi_{1}}^{eq} 
= \frac{g_{dof} M_{1}^{2} T}{2 \pi^{2}} K_{2}(M_{1}/T)$. 

The abundance of $Y_{\xi_{1}}=(n_{\xi_{1}^{-}}-n_{\xi_{1}^{+}})/s$, due to the decay and inverse decay of $\xi_{1}$ particles, satisfy the Boltzmann 
equation:
\begin{equation}
\frac{dY_{\xi_{1}}}{dz} = -\frac{\Gamma_{D}}{z H(z)}Y_{\xi_{1}} + \sum_{i=B-L,H}\frac{\Gamma_{ID}^{i}}{z H(z)} B_{i} Y_{i}
\end{equation}
where
\begin{equation}
\Gamma_{ID}^{i}=\Gamma_{D} (X_{\xi_{1}}^{eq}/X_{i}^{eq})\,\,\, {\rm and}\,\,\, B_{i}=\Gamma_{i}/\Gamma_{1}.
\end{equation}

Due to the conservation of hypercharge, the above Boltzmann equations satisfy the relation: $2 Y_{\xi}+\sum_{i}Y_{i}=0$\,\cite{Arina:2011cu}, 
where $i=B-L, H$. The evolution of $Y_i=n_i/s$, with $i=B-L, H$, is then given by,
\begin{equation}
\frac{dY_{i}}{dz} = 2 \Big\lbrace \frac{\Gamma_{D}}{z H(z)}\left[\epsilon_{i}(X_{\xi_{1}}-X_{\xi_{1}}^{eq})\right] + B_{i} 
\left( \frac{\Gamma_{D}}{z H(z)} Y_{\xi_{1}}- \frac{\Gamma_{ID}^{i}}{z H(z)} 2 Y_{i} \right)- \frac{\Gamma_{s}}{z H(z)} 
\frac{X_{\xi_{1}}^{eq}}{X_{L}^{eq}} 2 Y_{L} \Big\rbrace\,.
\end{equation}
The $\Gamma_{s}=\gamma_{s}/n_{\xi_{1}}^{eq}$ is the scattering rate of lepton number violating process, $LL \rightarrow HH$.
\begin{figure}[h!]
				\centering
				\includegraphics[width = 68mm]{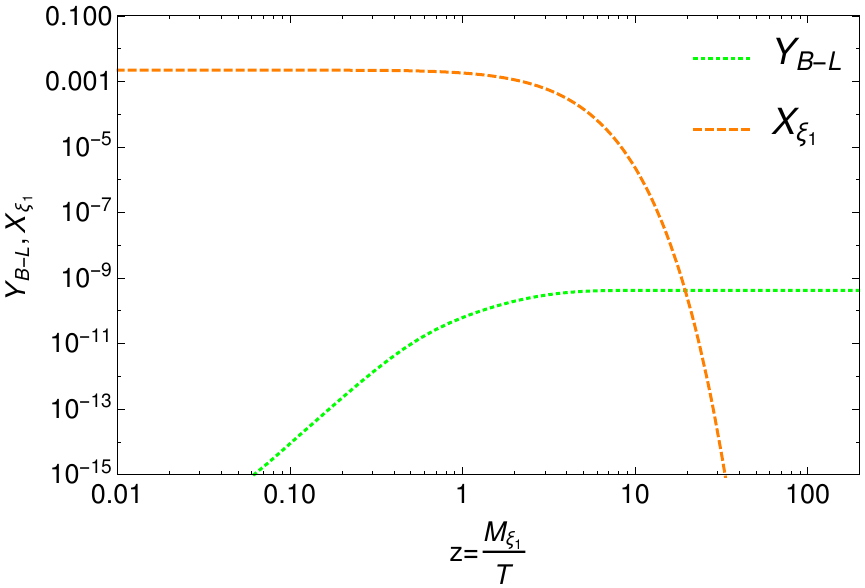}\,\,\,\,\,~~~~
				\includegraphics[width = 75mm]{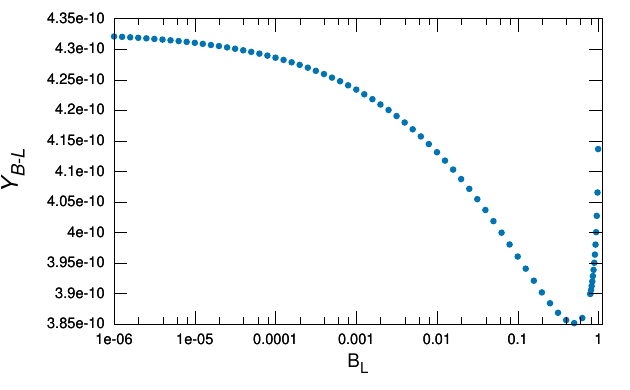}
				 \caption{\footnotesize{\textbf{Left)}.The abundance of $Y_{B-L}$ as a function of dimensionless variable $z=M_{1}/T$. We consider the parameter space $B_{L}=0.99$, $B_{H}=0.01$, $\epsilon_{L}\sim 10^{-6}$. \textbf{Right)}. The dependence of branching ratio, $B_{L}$, on $Y_{B-L}$ abundance.}}
                \label{BEq_Y_B-L}
\end{figure}
In the left panel of Fig.\ref{BEq_Y_B-L} we show the $B-L$ abundance $Y_{B-L}$ generated by the lightest triplet. We see that the $B-L$ yield agrees with the Planck observation~\cite{Aghanim:2018eyx}.  In the right panel of Fig.\ref{BEq_Y_B-L} we show the contribution of branching ratio $B_{L}$ to the abundance of $Y_{B-L}$. As the $B_{L}$ increases (which implies increasing dilution factor) the $Y_{B-L}$ decreases gradually until $B_{L} \approx 0.5$. For $B_L > 0.5$, $Y_{B-L}$ increases due to $\Gamma_1 \propto 1/\sqrt{B_{L}B_{H}}$ as given in Eq.\,\ref{Gamma_D}. 

Above electroweak phase transition a part of the $B-L$ asymmetry gets transferred to the baryon (B) asymmetry via the $B+L$ violating sphaleron 
transitions. The remaining $B-L$ asymmetry gets transferred to the dark sector by a higher dimensional operator~\cite{Feng:2012jn}: 
\begin{equation}
\mathcal {O}_8= \frac{1}{M_{asy}^{4}} \overline{\chi}^{2}(LH)^{2}\,.
\label{DLAsy_dim8}
\end{equation}
Note that this operator conserves $B-L$ and remains in thermal equilibrium above electroweak phase transition. 
As a result, the operator $\mathcal {O}_8$, redistributes the $B-L$ asymmetry between visible (comprising of SM fields) 
and dark (comprising of $\chi$) sectors. Since the DM candidate $\chi$ carries a Lepton number, $L=1$, we find that the 
lowest dimension operator allowed by the symmetry of our model that can transfer asymmetry from visible sector to dark sector is $\mathcal{O}_{8}
=\overline{\chi}^{2}(L H)^{2}/M_{asy}^{4}$.  Other possible lower dimensional operators, for example, dim-7, dim-6 and dim-5 allowed in this model, 
are not able to transfer the asymmetry from one sector to the other. So those operators will not alter the phenomenology of the asymmetric 
dark matter. See for instance ref.~\cite{asy_dm_zurek}.

The $B-L$ asymmetry of the visible sector gets transferred to a net baryon 
asymmetry via sphaleron transitions which conserve $B-L$ but violate $B+L$. The $B-L$ asymmetry 
of dark sector remains untouched and we see it as relics of DM of the present Universe. The symmetric component of 
the DM $\chi$ gets annihilated to SM fields through $\phi-H$ mixing which we study in the next section.

Note that the $B-L$ transfer operator ${\cal O}_8$ will decouple from the thermal plasma at different temperatures, 
depending on the value of $M_{asy}$. The decoupling temperature can be found by comparing the interaction rate of 
the operator with the Hubble expansion parameter. At the decoupling temperature $T_{D}$, the interaction rate 
$\Gamma_{\rm D}$ of the transfer operator $\mathcal{O}_8$ can be given as,
\begin{equation}
\Gamma_{\rm D} \simeq \left( \frac{T_D^4}{M_{\rm asy}^4}\right)^2 T_D\,.
\end{equation}
By comparing $\Gamma_{\rm D}$ with the Hubble expansion parameter $H (T)$ at a temperature $\rm T_{D}$ we get 
\begin{equation}\label{asy_constraint}
M_{asy}^{8} > M_{\rm Pl} T_D^{7}\,.
\end{equation}
This condition also implies that the processes allowed by the transfer operator will remain out of equilibrium 
below electroweak phase transition. Note that when $\chi$ mass is much smaller than the decoupling temperature 
$T_{D}$, then only the estimation of Eq.\,\ref{asy_constraint} holds.

Through out the above calculation we assume that the mass scale of the triplet scalar should be high 
enough so that the lepton asymmetry, generated by it, can be converted to the baryon asymmetry via 
sphaleron transitions before the latter processes decouple from thermal bath. That is the mass scale of 
triplet scalar $M_{\Delta}$ should be greater than the sphaleron decoupling temperature $T_{\rm sph} 
\approx 80 + 0.45 M_{h_{1}} = 136.33 > M_{W}$~\citep{Feng:2012jn,AristizabalSierra:2010mv,Strumia:2008cf,
Burnier:2005hp}. This implies that $M_{\Delta} \gg M_{W}$.  In our case,  we assume $T_D \gtrsim T_{\rm sph}$ 
i.e., the asymmetry transfer operator decouples before sphaleron processes decouple, which constrains 
$M_{\rm asy}$ using Eq.\,\ref{asy_constraint} to be $M_{\rm asy} > 0.9\times10^{4}{\rm GeV}$.

We assume that, the DM $\chi$ is in thermal contact with the visible sector until the $T_{\rm sph}> M_{W}$ 
through the higher dimensional operator $\mathcal{O}_{8}$. Therefore we get the number density of 
$\chi$ asymmetry (see Appendix ~\ref{appendix_asy}), which is nothing but the $B-L$ number density in dark sector, 
to be
\begin{equation}
n_\chi = (n_{\rm B-L})_{\rm dark} = -2\mu_{\chi} = \frac{58}{291}(n_{B-L})_{\rm vis}\,\,.
\label{a9}
\end{equation}

We also get the baryon asymmetry generated by the CP-violating out-of-equilibrium decay of $\xi_{1}$ in the visible 
sector as 
\begin{equation}
n_{\rm B}=\frac{30}{97} (n_{\rm B-L})_{\rm vis}\,\,.
\label{b-asymmetry}
\end{equation} 

Therefore, the total $B-L$ number density of the Universe generated by the CP-violating out of equilibrium decay 
of the scalar triplet $\xi_{1}$, is the sum of $n_{B-L}$ in the visible and dark sectors and is given by
\begin{eqnarray}
(n_{B-L})_{\rm total} &=& (n_{B-L})_{\rm vis} + (n_{B-L})_{\rm dark}\nonumber\\
&=& \frac{349}{291}(n_{B-L})_{\rm vis}.
\label{a11}
\end{eqnarray}

Comparing Eq.\,\ref{a11} with Eq.\,\ref{nb-ltotal_final} and using Eq.\,\ref{b-asymmetry}\,, we get the required 
CP asymmetry for observed lepton asymmetry $\epsilon_L = 141.23 (\eta/\kappa) (s/n_{\xi_{1}}^{eq} (T \rightarrow \infty ))$. 
Thus for $\kappa \sim 0.01$ we get $\epsilon_{L} \sim 10^{-6}$. Using Eq.\,\ref{a11} in Eq.\,\ref{b-asymmetry} and 
Eq.\,\ref{a9}, we get,
\begin{equation}
n_{\rm B}=\frac{90}{349} (n_{\rm B-L})_{\rm total} \,\,\,\,,\,\,\,\, n_\chi = \frac{58}{349}(n_{B-L})_{\rm total}
\label{nchi_nB_total}
\end{equation}

The present day ratio of DM relic density to baryon relic density, given by WMAP and the PLANCK data \cite{Hinshaw:2012aka, Aghanim:2018eyx}, 
is  $\Omega_{\rm DM} h^2/ \Omega_{\rm B} h^2 = 5.35 \pm 0.07$. This implies from Eq.\,\ref{nchi_nB_total} that,
\begin{equation}
M_{\chi} = \frac{\Omega_{\rm DM}h^2}{\Omega_{\rm B} h^2}(n_{B}/n_{\chi})\, M_{p}\approx 7.8\,\, {\rm GeV}
\end{equation}
where $M_{p}$ is the proton mass and $n_B/n_\chi=90/58$. Here onwards we discuss the constraints on the annihilation 
cross-section $\sigma(\bar{\chi} \chi \rightarrow SM)$ which depletes the symmetric component of 
the $\chi$-DM, for $M_{\chi} \approx 7.8 $ GeV.

\subsection{Depletion of symmetric component of the DM}
\label{symmetric_ann}
The symmetric component of $\chi$ can be efficiently annihilated through the $\phi-H$ mixing portal 
to the SM particles. Due to Breit-Wigner enhancement in the cross-section, when extra scalar mass 
$(M_{h_{2}})$ is twice of the DM mass, we get depletion of the symmetric component of the DM candidate. 
The annihilation cross-section of the process: $\overline{\chi}\chi \to \overline{f} f$, where $f$ is a SM 
fermion, is given by,
\begin{eqnarray}
\sigma_{\chi}  &=& \frac{\sqrt{s-4M_{f}^{2}}}{16 \pi s \sqrt{s}}  \nonumber\\
& \times & \frac{\lambda_{DM}^{2} \lambda_f^{2} \cos^{2}\gamma \sin^{2}\gamma}{\left[(s-M_{h_{1}}^{2})^{2}+\Gamma_{h_{1}}^{2}M_{h_{1}}^{2} \right]\left[(s-M_{h_{2}}^{2})^{2}+\Gamma_{h_{2}}^{2}M_{h_{2}}^{2} \right]} \nonumber\\
&\times & \lbrace \left[2 s-(M_{h_{1}}^{2}+M_{h_{2}}^{2}) \right]^{2} + \left[\Gamma_{h_{1}}M_{h_{1}}+\Gamma_{h_{2}}M_{h_{2}} \right]^{2} \rbrace \nonumber \\
& \times & \lbrace (s-2M_{\chi}^{2})(s-2M_{f}^{2})-2M_{f}^{2}(s-2M_{\chi}^{2}) \nonumber\\
& - & 2M_{\chi}^{2}(s-2M_{f}^{2})+4 M_{\chi}^{2} M_{f}^{2}\rbrace \,,
\label{scattering_cross_chi}
\end{eqnarray}
where $\sqrt{s}$ is the center of mass energy and $\lambda_f=M_f/v$ with $M_{f}$ being the mass of SM fermion $f$. The decay width of $h_1$ is given by:
\begin{equation}
\Gamma_{h_1}=\cos^2 \gamma \Gamma_{h_1}^{SM} + \sin^2 \gamma \Gamma_{h_1}^{\bar{\chi} \chi} + \Gamma_{h_1}^{h_2h_2}\,,
\label{gamma_h1}
\end{equation}
where $\Gamma^{SM}_{h_1}=4.2$ MeV and for $\Gamma_{h_1}^{\bar{\chi} \chi,\,h_{2}h_{2}},\, \Gamma_{h_{2}}$ refer to\,\cite{Narendra:2018vfw}. The thermal 
averaged annihilation cross-section of $\overline{\chi}\chi \to \overline{f} f$ is then given by~\cite{Gondolo&Gelmini}:
\begin{equation}
\langle \sigma_{\chi} v \rangle = \frac{1}{8 M_{\chi}^4 T K_2^2(M_\chi/T) } \times \int_{4 M_{\chi}^2}^\infty  \sigma_{\chi} (s-4 M_\chi^2) 
\sqrt{s} K_1(\sqrt{s}/T) ds  
\end{equation}
where $K_1 $ and $K_2$ are modified Bessel functions of first and second kind respectively and T is the temperature of thermal bath.  

\begin{figure}[h!]
				\centering
				\includegraphics[width = 80mm]{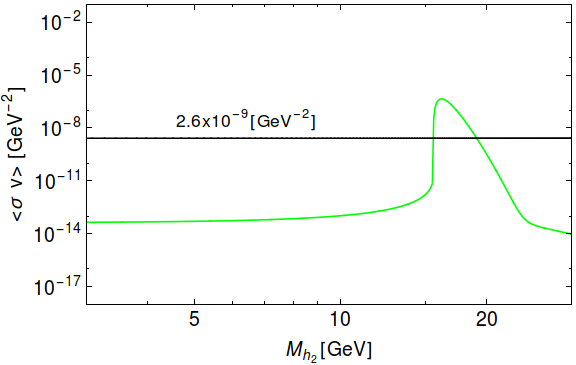}
				\includegraphics[width = 80mm]{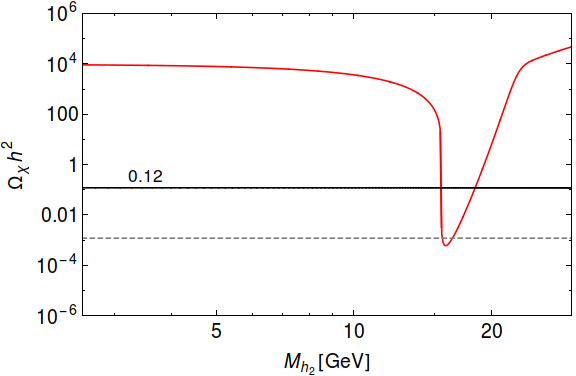}
					 \caption{\footnotesize{$\sigma v(\overline{\chi}\chi \rightarrow \overline{f}f)$ vs. $M_{h_{2}}$ (left plot) and the corresponding $\Omega_{\chi} h^{2}$ vs. $M_{h_{2}}$ (right plot) are shown for $\lambda_{\rm DM}=2\times 10^{-2}$ and $\sin \gamma=0.16$. In the right plot, the horizontal Black solid line and  Gray dashed line correspond to $\Omega_{\chi} h^{2}$ =0.12 ({\it i.e. 100\%} of DM relic) and $\Omega_{\chi} h^{2}$=0.0012 
({\it i.e. 1\%} of DM relic) respectively.}}
\label{depletion_chi}
\end{figure}

As we discussed at the end of previous Sec.\,\ref{trip_scalar_lep}, $\overline{\chi} \chi$ annihilates dominantly 
to the pairs of $\bar{b}b$, $\bar{\tau}\tau$ and $\bar{c}c$ particles. The unknown parameters which dominantly 
contribute to the annihilation cross-section in Eq.\,\ref{scattering_cross_chi} are the mass of $h_{2}$, {\it i.e.} 
$M_{h_{2}}$ and $\phi-H$ mixing, {\it i.e.} $\sin \gamma$ and the coupling of $h_{2}$ with $\chi$, {\it i.e.} $\lambda_{DM}$. 
All these parameters are constrained by invisible Higgs decay~\cite{Khachatryan:2016whc}, 
relic abundance of DM measured by Planck~\cite{Aghanim:2018eyx} and WMAP~\cite{Hinshaw:2012aka}, 
and spin-independent direct detection cross-section at XENON100~\cite{Aprile:2012nq}, LUX~\cite{Akerib:2016vxi} 
and XENON1T~\cite{Aprile:2015uzo}, \cite{Xenon1t_2018}\,, and the measurement of signal strength of the Higgs particle 
at LHC~\cite{CMS:2018lkl, Khachatryan:2016vau}. As discussed in Sec.\,\ref{Model}, we set $\sin \gamma=0.16$. Moreover, we 
fixed $\lambda_{DM}= 2 \times 10^{-2}$ to estimate the value of cross-section $\langle \sigma_{\chi} v \rangle$. We then plotted 
the thermal averaged annihilation cross-section $\langle \sigma_\chi | v\rangle$ as a function of $M_{h_{2}}$ in Fig.\,\ref{depletion_chi}\,. 
Here we fixed $M_\chi/T=20$ as we expect the maximum annihilation of $\overline{\chi}\chi \to \overline{f} f$ occurs at a temperature 
$T=M_\chi/20$. From the left plot of Fig.\,\ref{depletion_chi}, we see that in most of the parameter space $\langle \sigma_\chi |v| \rangle$ 
is less than the $\langle \sigma_\chi |v| \rangle_F=2.6\times 10^{-9}/{\rm GeV^{2}}$, but near the resonance region, it satisfy 
the condition $\langle \sigma_\chi |v| \rangle > \langle \sigma_\chi |v|\rangle_F$. Note that a large cross-section is 
required to completely annihilate the symmetric component of the DM and it can be achieved near the resonance, where 
the mass of $h_{2}$ is nearly twice of the DM mass. Far from the resonance we have $\langle \sigma_\chi |v| \rangle < \langle \sigma_\chi |v| 
\rangle_F$. Therefore, we get an over abundance of DM. In the right plot of Fig. \ref{depletion_chi}, we have shown relics of 
symmetric component of $\chi$ as a function of $M_{h_2}$. We see that the symmetric component of DM can be annihilated significantly  
only when $M_{h_{2}}$ varies in the range $15.6 - 16.7$ GeV. This shows that our result crucially depends on $M_{h_2}$. We will come back to this point 
while giving a summary plot in Fig.\ref{lambdaDM_vs_singm}.

\subsection{Higgs signal strength}\label{signal_strength}
The Higgs signal strength is defined for a particular channel $h_{1}\rightarrow x x$ as
\begin{eqnarray}
\mu_{h_1 \to xx} &=& \frac{\sigma_{h_1}}{\sigma_{h_1}^{\rm SM}} \frac{{\rm Br}_{h_1 \to xx} }{{\rm Br}^{\rm SM}_{h_1 \to xx}} \nonumber\\
&=& \frac{\cos^4 \gamma \Gamma_{h_1}^{SM} }{ \Gamma_{h_1}}\,,
\end{eqnarray}
where $\Gamma_{h_1}$ is given in Eq.~\ref{gamma_h1}. The $\sigma_{h_1}$ and $\sigma_{h_1}^{\rm SM}$ are Higgs production cross section in BSM model and SM respectively.
Higgs signal strength measurement at LHC can constrain $\phi-H$ mixing in our model. Currently the combined Higgs signal strength measured value is $\mu=1.17 \pm 0.1$~\cite{CMS:2018lkl,Khachatryan:2016vau}. We have taken $2\sigma$ and $3\sigma$ deviations from the best fit value to constrain our model parameters, the corresponding contour lines are shown in Fig.\,\ref{lambdaDM_vs_singm}.
 
\subsection{Constraints from invisible Higgs decay}\label{invisible}
This model allows SM-like Higgs $h_{1}$ to decay via invisible channels through $\phi-H$ mixing: $h_{1} \rightarrow h_{2}h_{2},\, h_{1} \rightarrow \overline{\chi}\chi$. The branching ratio for the invisible Higgs decay can be 
defined as: 
\begin{equation}
Br_{\rm inv} = \frac{ \sin^2 \gamma \Gamma_{h_1}^{\bar{\chi} \chi} + [{\rm Br}(h_2 \to \bar{\chi} \chi )] \Gamma_{h_1}^{h_2h_2}} {\cos^2 \gamma \Gamma_{h_1}^{SM} + \sin^2 \gamma \Gamma_{h_1}^{\bar{\chi} \chi} + \Gamma_{h_1}^{h_2h_2}}
\label{invisible_higgs_decay}
\end{equation}
 Note that LHC gives an upper bound to the invisible Higgs decay as $Br_{\rm inv} \leq 24\% $~\cite{Khachatryan:2016whc}. This bound on Higgs invisible decay width constraint the $\lambda_{\rm DM}$ and $\sin \gamma$ in our model as shown in Fig.\,\ref{lambdaDM_vs_singm}.
 
\subsection{Constraints from direct detection of DM}\label{dd}
In our setup, the $\phi-H$ scalar mixing allows the DM $\chi$ to scatter off the target nucleus which can be checked at terrestrial laboratories. The spin 
independent DM-nucleon elastic scattering cross-section per nucleon can be written as \cite{Goodman:1984dc,Ellis:2008hf,Akrami:2010dn,Bhattacharya:2015qpa,Patra:2016shz,Bhattacharya:2017sml,Ellis:2000ds} ,
 \begin{equation}
 \sigma^{SI} = \frac{\mu_{r}^{2}}{\pi A^{2}}[Z f_{p}+(A-Z)f_{n}]^{2}
 \label{DD1}
 \end{equation}
where the A and Z are the mass and atomic numbers of the target nucleus and  $\mu_{r}$ is the reduced mass and is given by $\mu_{r} = M_{\chi}m_{N}/(M_{\chi}+m_{N})$, where $m_{N}$ is the mass of the nucleon (proton or neutron). In Eq.\,\ref{DD1} $f_{p}$ and $f_{n}$ are the effective interaction strengths of DM with proton and neutron of the target nucleus and are given by:
	\begin{equation}
	f_{p,n}=\sum\limits_{q=u,d,s} f_{T_{q}}^{p,n} \alpha_{q}\frac{m_{p,n}}{m_{q}} + \frac{2}{27} f_{TG}^{p,n}\sum\limits_{q=c,t,b}\alpha_{q} 
\frac{m_{p,n}}{m_{q}}\,,
\label{DD2}
	\end{equation}
where 
\begin{equation}
 \alpha_{q} = \lambda_{DM} \left( \frac{m_{q}}{v}\right) \left[\frac{1}{M_{h_{2}}^{2}}-\frac{1}{M_{h_{1}}^{2}}\right] \sin\gamma \cos\gamma\,.
  \label{DD4_chi}
 \end{equation}
In Eq.\,\ref{DD2}, the $f_{T_{q}}^{p,n}$ are the quark mass fractions inside the nucleons, defined as $m_{N}f_{T_{q}}^{p,n}\equiv \braket{N|m_{q}\bar{q}q|N}$ and their values are $f^{(p)}_{Tu}= 0.020 \pm 0.004, f^{(p)}_{Td}= 0.026 \pm 0.005, f^{(p)}_{Ts}= 0.118 \pm 0.062, f^{(n)}_{Tu}=0.014 \pm 0.003, f^{(n)}_{Td}= 0.036 \pm 0.008, f(n)_{Ts}= 0.118 \pm 0.062$\,\cite{Ellis:2000ds,Narendra:2018vfw}. The coupling strength of DM with the gluons in the target nuclei is parameterized as
	\begin{equation}
	f_{TG}^{p,n}=1-\sum\limits_{q=u,d,s} f_{T_{q}}^{p,n}\,.
	 \label{DD3}
	\end{equation}

\begin{figure}[h!]
				\centering
				\includegraphics[width = 80mm]{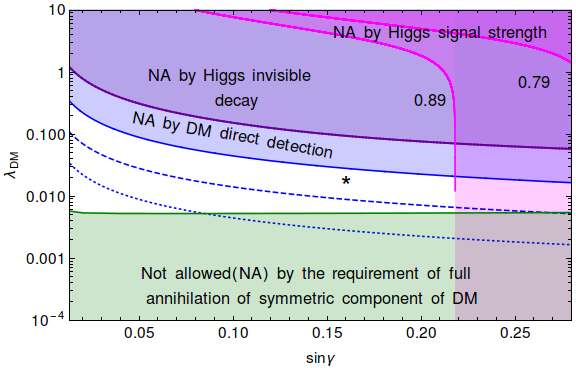}
				\includegraphics[width = 80mm]{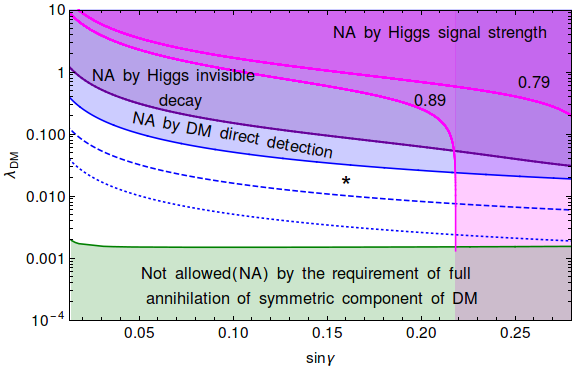}
				 \caption{\footnotesize{$\lambda_{DM}$ versus $\sin \gamma$ parameter space. The region above the Blue, Purple and Pink lines are disallowed by the spin independent dark matter nucleon cross-section of $10^{-43} {\rm cm}^2$ at XENON1T\cite{Xenon1t_2018} for DM mass 7.8 GeV, the invisible Higgs decay, {\it i.e.} $Br_{\rm inv} \geq 24 \%$ and the Higgs signal strength for $\mu=0.79,\,0.89$, respectively. The region below the bottom green line is not allowed $\langle\sigma v \rangle < 2.6 \times 10^{-9}/{\rm GeV}^2$ because it does not satisfy the full annihilation of the symmetric component of the DM.  The Left and Right panels correspond to $M_{h_{2}}$=15.6 GeV and 16.7 GeV respectively. In both panels we fixed $M_\chi =7.8$ GeV. The star corresponds to a point $\lambda_{DM}$=$2\times 10^{-2}$ and $\sin \gamma=0.16$. The Blue dashed and dotted lines correspond to DM-nucleon cross-sections of $10^{-44} {\rm cm}^2$  and $10^{-45} {\rm cm}^2$ respectively for $M_\chi=7.8$ GeV.}}
\label{lambdaDM_vs_singm}
\end{figure}

In Fig.~\ref{lambdaDM_vs_singm} we combined all the constraints coming from the invisible Higgs decay, Higgs signal strength, direct detection of DM at 
XENON1T~\cite{Xenon1t_2018} and the relic abundance of dark matter in the plane of $\lambda_{\rm DM}$ versus $\sin \gamma$. In the left-panel of Fig.~\ref{lambdaDM_vs_singm} we have taken $M_{h_{2}}$=15.6 GeV, while that in the right panel $M_{h_{2}}$=16.7 GeV. The Pink shaded region shows the constraint 
from Higgs signal strength measurement ($\mu=1.17 \pm 0.1$)~\cite{CMS:2018lkl,Khachatryan:2016vau}, the Purple region shows the constraint from invisible 
Higgs decay (${\it i.e.,}$ $Br_{\rm inv} \geq 24 \%$)~\cite{Khachatryan:2016whc}, while the Blue region is disallowed by the spin independent direct 
detection cross-section at XENON1T where we have used DM-nucleon cross-section $10^{-43} {\rm cm}^2$ for a DM mass 7.8 GeV, and the Green region is 
disallowed by the constraint coming from the relic abundance of DM (${\it i.e.,}$ $ \langle \sigma v (\bar{\chi} \chi \, \to \bar{f} f)\rangle 
< 2.6\times 10^{-9}/{\rm GeV}^2$). Finally, we end up with a white region of allowed parameter space in the plane of $\lambda_{\rm DM}$ versus 
$\sin \gamma$. In the white region, the projected Blue dashed and dotted lines, which correspond to a DM-nucleon cross-section of $10^{-44} {\rm cm}^2$ 
and $10^{-45} {\rm cm}^2$ respectively for a DM mass $M_\chi=7.8$ GeV, can be probed by future data.  

\section{Signature of light Higgs}\label{collider-signature}
In our model, apart from the SM Higgs boson $h_1$, there exist an extra scalar particle $h_2$, which plays an important role in the 
annihilation of symmetric component of the dark matter. As discussed before, the mass of $h_2$ is required to be  around $16\, \rm GeV$ to 
annihilate the symmetric component of the dark matter. Here we briefly discuss the collider signature of a light Higgs pertinent to this model.

At LHC, the main production channel of $h_2$ is through the decay of SM Higgs $h_1$, {\it i.e.} $h_{1}\rightarrow h_{2}h_{2}$. The branching 
fraction of $h_1 \to h_2 h_2$ is about $3.58\%$ (for a typical set of values: $\sin\gamma = 0.16$, $\lambda_{\rm DM}=0.01$, $\lambda_H=0.129$, 
$\lambda_{\rm H\phi}=0.01$, $M_{h_2}=16.42$ GeV). The subsequent decay of $h_2$ to SM particles can be studied at collider. We note that the main 
decay modes of $h_2$ in this model are $\rm Br(h_{2}\rightarrow b\bar{b})\sim 74.15\%$, $\rm Br(h_{2}\rightarrow \tau^{+}\tau^{-})\sim 6.51\%$, 
$Br(h_{2}\rightarrow c\bar{c})\sim 10.43\%$, $Br(h_{2}\rightarrow \mu^{+}\mu^{-})\sim  0.0741\%$ and $Br(h_{2}\rightarrow \bar{\chi}\bar{\chi})
\sim  8.90\%$. At LHC the main production channel of the SM Higgs is gluon-fusion and the corresponding 
cross-section for $gg\rightarrow h_{1} \rightarrow h_{2}h_{2}$ is given by $1.78\, \rm pb$\,\cite{Dittmaier:2011ti} at c.m.energy $14\,\rm TeV$. 
Depending on the decay mode the final state will be vary, such as $b\bar{b}b\bar{b}$, $b\bar{b}\tau^{+}\tau^{-}$, 
$\tau^{+}\tau^{-}\tau^{+}\tau^{-}$, $\tau^{+}\tau^{-}\mu^{+}\mu^{-}$, {\it etc.}, for the search of $h_2$ at LHC. All of 
these channels have either large backgrounds, which are mostly dominated by the QCD, or very small cross-section. So at LHC, it is very 
difficult to search this light Higgs boson. Despite this a large number of searches have been performed in the last years. A most recent search result 
is given in \cite{ATLAS:2021ypo}]. For $M_{h_{2}}=16\rm\, GeV$ the bound on $Br(h_{1}\rightarrow h_{2}h_{2}\rightarrow b\bar{b}\mu^{+}\mu^{-})\lesssim 10^{-4}$,
which is compatible with the branching ratios mentioned above. There are other existing searches by CMS and ATLAS for various final states in the 
different mass range of $h_2$. For example, 
$4\mu$ in the final state with $M_{h_{2}}$ varying in the range $1- 15$ GeV~\cite{CMS:2018jid,ATLAS:2018coo}, $2\mu 2\tau$ in the final state with $M_{h_{2}}$ 
varying in the range $3.6-21$ GeV~\cite{CMS:2020ffa}, two muons and two tracks in the final state with $M_{h_{2}}$ varying in the range 
$4-15$ GeV~\cite{CMS:2019spf}, two bottom quarks and two tau leptons in the final state with $M_{h_{2}}$ varying in the range $15-60$ GeV~\cite{CMS:2018zvv}, 
four bottoms in the final state with $M_{h_{2}}$ varying in the range $15-30$ GeV~\cite{ATLAS:2020ahi}, $4\gamma$ in the final state with $M_{h_{2}}$ varying 
in the range $10-62$ GeV\cite{ATLAS:2015rsn}, $\gamma\gamma j j$ in the final state with $M_{h_{2}}$ varying in the range: $20-60$ GeV~\cite{ATLAS:2018jnf}, etc. 
However, none of the searches are fruitful yet.

The signature of the light scalar $h_2$ can also be studied with a better precession at leptonic colliders, such as the International Linear Collider (ILC), 
which is proposed to run at the center of mass energies 500 GeV and 1 TeV. The main production channel of the light Higgs $h_2$ at ILC is via the process 
$e^{+}e^{−}\rightarrow Z h_{1}$ and subsquesnt decays of $h_{1} \to h_{2} h_{2}$~\cite{Drechsel:2018mgd}. Alternatively the direct production of $h_2$ can 
happen through the process:  $e^{+} e^{−} \to  Z h_{2}$~\cite{Wang:2020lkq,Drechsel:2018mgd}. Subsequently these particles decay to SM particles and pave a 
way to detect $h_2$. In fact, the analysis of ref. \cite{Wang:2020lkq} shows that ILC can even be able to detect a 10 GeV $h_2$ with a $\phi-H$ mixing of 
order $10^{-2}$.

\section{Conclusion}\label{conclusion}
In this paper, we studied a simultaneous explanation of visible and dark matter in a type-II seesaw 
scenario, thus explaining why DM to baryon ratio in the present Universe is about a factor of five. We extended the 
standard model with two triplet scalars $\Delta_{i}, (i=1, 2)$ and a singlet leptonic Dirac 
fermion $\chi$. The particle $\chi$ is charged under an extended global symmetry $U(1)_{D}$, which is softly broken by 
dimension-8 operator $(\bar{\chi}LH)^2/M_{asy}^{4}$ to a remnant $Z_{2}$ symmetry under which $\chi$ is odd, while all 
other particles are even. As a result $\chi$ served as a stable DM candidate.

The lightest triplet scalar creates a net $B-L$ asymmetry in the early Universe via its CP violating out-of-equilibrium 
decay to SM leptons and Higgs. The created $B-L$ asymmetry is then transferred to the dark sector via the dimension-8 
operator, which conserves $B-L$ number. The transfer of $B-L$ asymmetry is active until the dimension-8 operator decouples 
from the thermal bath. As a result, there is a proportionality arises between the net $B-L$ asymmetry in the visible and 
the dark sector. Note that the dimension-8 operator decouples from the thermal bath before the decoupling of sphaleron 
processes. As a result, the $B-L$ asymmetry in the visible sector gets converted to a net B-asymmetry through the 
sphaleron transitions, while the $B-L$ asymmetry of dark sector remains untouched which we see today as relics of 
asymmetric $\chi$-particles. A singlet scalar $\phi$ is then introduced to deplete the symmetric component of the DM $\chi$ 
through its mixing with the SM-Higgs. We found that nearly 100\% depletion of the symmetric component of $\chi$-DM is 
possible within a narrow range of singlet scalar mass, namely (15.6 - 16.7) GeV, where DM mass is about half of 
singlet scalar mass, i.e., $M_\chi \sim 8$ GeV. By considering the constraints from invisible Higgs 
decay, Higgs signal strength, null detection of DM at XENON1T and relic abundance of DM, we showed the allowed region 
of parameters in the plane of DM coupling $\lambda_{DM}$ and $H-\phi$ scalar mixing: $\sin \gamma$. 

After electroweak phase transition, the scalar triplet $\Delta$ acquires an induced vacuum expectation value 
$\langle \Delta \rangle$. As a result the sub-eV neutrino masses could be explained through the $\Delta LL$ coupling, 
assuming $\mu \sim M_\Delta \sim  10^{14}$ GeV.

\section*{Appendix}
\subsection{Asymmetry transfer from visible to dark sector}\label{appendix_asy}
The asymmetry in the equilibrium number densities of particle $n_{i}$ over antiparticle $\overline{n}_{i}$ can be written as
\begin{equation}
n_{i}-\overline{n}_{i} = \frac{g_{i}}{2 \pi^{2}} \int_{0}^{\infty}dq ~ q^{2} \left[ \frac{1}{e^{\frac{E_{i}(q)-\mu_{i}}{T}}\pm 1}-\frac{1}{e^{\frac{E_{i}(q)+\mu_{i}}{T}}\pm 1} \right]\,,
\end{equation}
where the $g_{i}$ is the internal degrees of freedom of the particle species $i$. In the above equation $E_{i}$ and $q_{i}$ represent 
the energy and momentum of the particle species $i$. In the approximation of a weakly interacting plasma, where $\beta \mu_{i} \ll 1$, $\beta \equiv 1/T$ (for further detailed discussion visit~\cite{Feng:2012jn},\cite{Kolb:1990vq}) we get 
\begin{eqnarray}
n_{i}-\overline{n}_{i} & \sim & \frac{g_{i}T^{3}}{6} \times [ 2\beta\mu_{i}+\mathcal{O}\left((\beta \mu_{i})^{3}\right)~~~~ {\rm for \,\,bosons} \nonumber\\
& \sim & \frac{g_{i}T^{3}}{6} \times [ \beta \mu_{i}+\mathcal{O}\left((\beta \mu_{i})^{3}\right) ~~~~~{\rm for \,\,fermions}.
\end{eqnarray}

In our model, the asymmetry transfer operator is given by ${\cal O}_8= \frac{1}{M_{asy}^{4}} \overline{\chi}^{2}(LH)^{2}$. Depending 
on the value of the $M_{\rm asy}$ the operator will decouple from thermal plasma at different temperatures. Since the $B-L$ asymmetry 
generated by the decay of scalar triplet is required to be transferred to the dark sector via ${\cal O}_8$ operator, we assume the decoupling temperature $T_D$ of the latter to be $T_{t} > T_{D} > T_{W}$, where $T_{t}$ is the temperature of thermal bath when the top quark decouples and $T_{W}$ is the temperature when the $W$ boson decouples from the thermal plasma. In this case the effective Lagrangian for Yukawa coupling is given by:
\begin{equation}
\mathcal{L}_{Yukawa} = g_{e_{i}}^{k}\bar{e}_{iL}h^{k}e_{iR}+g_{u_{i}}^{k} \bar{u}_{iL}h^{k}u_{iR}+g_{d_{i}}^{k}\bar{d}_{iL}h^{k}d_{iR} +h.c
\label{b1}
\end{equation}
where $k=1,\,2$ for two Higgses $h_{1}$ and $h_{2}$. As Higgs field is real so the above Lagrangian gives the following chemical equilibrium condition:
\begin{equation}
0=\mu_{h}=\mu_{u_{L}}-\mu_{u_{R}}=\mu_{d_{L}}-\mu_{d_{R}}=\mu_{e_{L}}-\mu_{e_{R}}.
\label{b2}
\end{equation}
The charged current interaction part of the SM Lagrangian after electroweak symmetry breaking is given by:
\begin{equation}
\mathcal{L}_{int}^{(W)}=gW_{\mu}^{+}\bar{u}_{L}\gamma^{\mu} d_{L} + gW_{\mu}^{+}e_{L}\gamma^{\mu}\bar{\nu}_{eL}\,.
\label{b3}
\end{equation}
The above equation implies that the charged current interactions remain in thermal equilibrium until W-boson decouples 
from thermal bath. As a result we get the following chemical potential constraints:
\begin{equation}
\mu_{W}=\mu_{u_{L}}-\mu_{d_{L}},
\label{b4}
\end{equation}
and
\begin{equation}
\mu_{W}=\mu_{\nu}-\mu_{e_{L}}.
\label{b5}
\end{equation}
The electroweak sphalerons remain in thermal equilibrium until a temperature $T_{\rm sph} \gtrsim  T_{W}$. As a result 
we get a constraint:
\begin{equation}
\mu_{u_{L}}+2\mu_{d_{L}}+\mu_{\nu}=0.
\label{b6}
\end{equation}

At a temperature below electroweak phase transition, the electric charge neutrality of the Universe holds.  However, 
at the epoch: $T_{t} > T_{D} > T_{W}$, the  top  quark  is  already  decoupled from the thermal plasma and hence 
does not take part in the charge neutrality condition. Therefore, we get 
\begin{equation}
Q=4(\mu_{u_{L}}+\mu_{u_{R}})+6\mu_{W}-3(\mu_{d_{L}}+\mu_{d_{R}}+\mu_{e_{L}}+\mu_{e_{R}})=0.
\label{c1}
\end{equation}
Now using the Eqs.\,\ref{b2}-\ref{c1}, the baryon and lepton number density $n_{B}$ and $n_{L}$ can be written as,
\begin{equation}
n_{B}=-\frac{90}{19}\mu_{\nu}
\label{c2}
\end{equation}
and 
\begin{equation}
 n_{L}=\frac{201}{19}\mu_{\nu}\,,
\label{c3}
\end{equation}
where we have dropped the common factor $g T^3\beta/6$ as we are interested in ratio of 
densities, rather than their individual values. The net $B-L$ asymmetry in the visible sector is thus given by:
\begin{equation}
(n_{B-L})_{\rm vis}=-\frac{291}{19}\mu_{\nu}\,.
\label{c4}
\end{equation}
After sphaleron processes decouple at $T_{\rm sph}$, the baryon and lepton number densities would be conserved separately. 
As a result Eqs.\,\ref{b2} -~\ref{c4} would remain valid at 
$T_{\rm sph} > T_{D} > M_{W}$. Once the sphaleron processes decouple, the ratio of $n_{B}/n_{B-L}$ would be frozen. As 
a result from Eqs.\,\ref{c2}  and \ref{c4}, it can be written as,
\begin{equation}
\frac{n_{B_{\rm final}}}{(n_{B-L})_{\rm vis}}=\frac{n_{B}}{(n_{B-L})_{\rm vis}}=\frac{30}{97}=0.31
\label{c5}
\end{equation}
\begin{equation}
n_{B_{\rm final}}=0.31(n_{B-L})_{\rm vis}.
\label{c6}
\end{equation}
As the asymmetry transfer operator ${\cal O}_8$ is active down to W-boson decoupling temperature, a part of the $B-L$ 
asymmetry gets transferred from visible sector to the dark sector. Therefore, $(n_{B-L})_{\rm vis}$ is no longer equal 
to the $(n_{B-L})_{\rm total}$. 

The equilibration of ${\cal O}_8$ operator gives the constraint: 
\begin{equation}
\mu_{\chi} = \mu_{\nu} \,.
\label{c8}
\end{equation}
As a result the number density of dark matter $\chi$, which is nothing but the $B-L$ number density of dark sector, 
is given by:
\begin{equation}
n_{\chi}=-2\mu_{\chi}=\frac{58}{291}(n_{B-L})_{\rm vis} = (n_{B-L})_{\rm dark}
\label{c8}
\end{equation}
We use the $B-L$ number density of $\chi$ to calculate its mass in Sec.\,\ref{lep_dm}. 

\subsection{A viable origin of dimension-8 operator}\label{origin_dim8}
We now discuss a viable origin of the dimension-8 operator, ${\cal O}_8=\frac{\bar{\chi}^{2}(LH)^{2}}{M_{asy}^{4}}$, which conserves 
$B-L$ symmetry but breaks $U(1)_D$ global symmetry explicitly to a remnant $Z_2$ symmetry under which the DM $\chi$ is odd. Apart from 
the singlet fermion $\chi$, we add a relatively heavy scalar doublet (under $SU(2)_L$) $\eta$ to the dark sector. We assume that $\eta$ 
is odd under $Z_2$ symmetry and possesses same charge under $U(1)_D$ as that of $\chi$. As a result the relevant Lagrangian, which can 
give rise to the required dimension-8 operator, can be given as: 
\begin{equation}
\mathcal{L} \supset f ~\overline{\chi} L \eta + \lambda_{_{\eta H}}(\eta^{\dagger} H)^{2} + {\rm h.c.}\,,
\end{equation}
where $\eta=(\eta_{1}+ i \eta_{2})/\sqrt{2}$.
\begin{figure}[h!]
				\centering
				\includegraphics[width = 45mm]{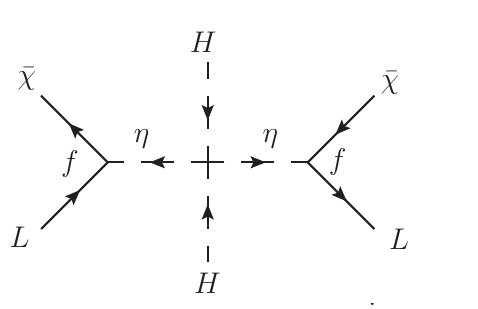}
               \caption{\footnotesize{Feynman diagram of the dimension-8 operator.}}
              \label{asydm_cartoon}
\end{figure}
The Feynman diagram, which in the effective theory can give rise to a dimension-8 operator, is shown in Fig.\ref{asydm_cartoon}. We assume that $\eta$ 
is heavy and doesn't acquire a vev in order to preserve the remnant $Z_2$ symmetry of the dark sector. As a result, integrating $\eta$, we get the 
required dimension-8 operator:
\begin{equation}
\frac{f^{2} \lambda_{_{\eta H}} \overline{\chi} L H H \overline{\chi} L}{M_{\eta}^{4}} \equiv \frac{\overline{\chi}^{2}(L H)^{2}}{M_{asy}^{4}}\,,
\end{equation}
where $M_{asy}^{4}=M_{\eta}^{4}/(f^{2} \lambda_{_{\eta H}})$.

With the introduction of $\eta$ the additional terms in potential are
\begin{equation}
M_{\eta}^{2} (\eta^{\dagger} \eta) + \lambda_{\eta}(\eta^{\dagger} \eta)^{2} + \lambda_{_{\eta H}} (\eta^{\dagger}\eta) (H^{\dagger}H) +\lambda_{_{\eta H}} (\eta^{\dagger} H)^{2} + h.c. .
\end{equation}
The last term $\lambda_{_{\eta H}} (\eta^{\dagger} H)^{2} + h.c.$ in the potential breaks the $U(1)_{D}$ symmetry. This term will generate a mass splitting between $\eta_{1}$ and $\eta_{2}$. The breaking is related to the mass splitting and does not have any impact on other part of the model.
\section*{Acknowledgements}
The computations was supported in part by the SAMKHYA: High Performance Computing Facility provided by Institute of Physics, Bhubaneswar.
{}
\end{document}